\documentclass[pra,aps,twocolumn,epsfig,nofootinbib]{revtex4}
\usepackage{amsmath}
\usepackage{amssymb}
\usepackage[dvips]{graphicx}%\usepackage{graphicx}% Include figure files
\usepackage{dcolumn}% Align table columns on decimal point
\usepackage{bm}% bold math
\usepackage[colorlinks=false,dvipdfm]{hyperref} % citecolor=blue 文件内部引用
\usepackage{setspace}
\usepackage{amsfonts}
\usepackage{rotating}
\usepackage{makeidx}
\usepackage{amsthm}

\newcommand{\tr}{\mbox{Tr}}

\newcommand{\Diag}{\mbox{Diag}}
\newcommand{\sgn}{\mbox{sgn}}

\begin{document}
%\begin{CJK*}{GB}{}

\title{Local-hidden-state models for Bell diagonal states and beyond \footnote{Phys. Rev. A. 99, 062314 (2019)}}

\author{Fu-Lin Zhang}
\email[Corresponding author: ]{flzhang@tju.edu.cn}
\affiliation{Department of Physics, School of Science, Tianjin University, Tianjin 300072, China}

\author{Yuan-Yuan Zhang}
\affiliation{Department of Physics, School of Science, Tianjin University, Tianjin 300072, China}

\date{\today}

\begin{abstract}
For a bipartite entangled state shared by two observers, Alice and Bob, Alice can affect the post-measured states left to Bob by choosing different measurements on her half.
Alice can convince Bob that she has such an ability if and only if the unnormalized postmeasured states cannot be described by a local-hidden-state (LHS) model.
In this case,  the state is termed steerable from Alice to Bob.
By converting the problem to construct LHS models for two-qubit Bell diagonal states to the one for  Werner states, we obtain the optimal models given by Jevtic \textit{et al.} [J. Opt. Soc. Am. B \textbf{32}, A40 (2015)],  which are developed by using the steering ellipsoid formalism.
Such conversion also enables us to derive a sufficient criterion for unsteerability of any two-qubit state.

\end{abstract}

%\pacs{03.67.Mn 03.65.Ta 42.50.Dv

%\pacs{03.65.-w; 02.20.-a; 21.10.sf; 31.30.jx}
%\keywords{}

\maketitle
%\end{CJK*}

 \section{Introduction}\label{Intro}

 Many concepts have been presented to describe the nonclassical correlations in composite quantum  systems \cite{Book,RevModPhys.81.865,RMP2012Vedral,RMP2014bell,JPA2014LHV}.
The definitions of  these correlations rely on the division between the quantum and classical worlds \cite{RMP2012Vedral}.
Entanglement \cite{RevModPhys.81.865} is the most prominent of these correlations, which exists in the quantum systems whose states cannot be expressed as a mixture of product states.
 The violation of the local-hidden-variable (LHV) model by the outcomes of  local measurements demonstrates the Bell-nonlocality \cite{RMP2014bell,JPA2014LHV}.
 Entanglement is a necessary condition of  Bell nonlocality, as any separable (unentangled) state can be modeled by a LHV theory.
 However, the sufficiency of this statement does not hold up.
 This result was found by Werner \cite{Werner1989}, who constructed an explicit LHV model of a family of highly symmetric mixed entangled states, known today as the Werner states.
 That is, Bell nonlocality is a stronger correlation than entanglement.

Einstein-Podolsky-Rosen (EPR) steering lies between Bell nonlocality and entanglement \cite{PRL2007Steering}.
The term \textit{steering} was introduced by Schr\"{o}dinger \cite{steer1935}.
It means that when observers Alice and Bob share an entangled state, Alice can prepare Bob's system into different states by choosing her local measurement.
The singlet state of two qubits  can serves as the simplest example.
It is given by
\begin{eqnarray}\label{SigState}
|\Phi^-\rangle_{AB}=\frac{1}{\sqrt{2}}\bigr(|0\rangle|1\rangle-|1\rangle|0\rangle\bigr)=\frac{1}{\sqrt{2}}\bigr(|\!+\!\rangle|\!-\!\rangle-|\!-\!\rangle|\!+\!\rangle\bigr),
\end{eqnarray}
where $|0\rangle$ and $|1\rangle $ are the eigenstates of the third Pauli operator $\sigma_z$, and $|\pm\rangle=(|0\rangle\pm|1\rangle)/\sqrt{2}$ are the ones of the first Pauli operator  $\sigma_x$,  satisfying $\sigma_z |0\rangle=|0\rangle $ and $\sigma_z |1\rangle=-|1\rangle$, and $\sigma_x |\pm\rangle=\pm |\pm\rangle$, respectively.
Alice can project Bob's system into one of the states $|0\rangle$ and $|1\rangle $, or of $|+\rangle$ and $|-\rangle$, by measuring on $\sigma_z$ or  $\sigma_x$.

The operational definition of EPR steering was provided by Wiseman \textit{et al.} \cite{PRL2007Steering}.
An entangled state $\rho_{AB}$, shared by Alice and Bob, is said to have EPR steering from Alice to Bob, when it can be used to demonstrate Alice's ability of steering.
This is equivalent to the set of unnormalized postmeasured states in Bob's hands,  often referred to as an assemblage, which cannot by described by a local-hidden-state (LHS) model.
Intuitively, the LHS model provides a simulation of Alice's outcomes and  the postmeasured states, by using a preparation strategy of Bob's state in which no entanglement is involved.

The applications of  EPR steering have been explored in quantum information processing, such as quantum key distribution \cite{PRA2012OneSDIQKD} and randomness generation \cite{JPA2014quantum}.
Experimental investigations \cite{NT2010Exp,wittmann2012loophole} and application in detecting entanglement \cite{he2011einstein} have also been reported.

However, the explicit construction of LHS models is an extremely difficult problem even for the simplest case of two qubits.
 Wiseman \textit{et al.} \cite{PRL2007Steering} pointed out that the LHV model in Werner's seminal work \cite{Werner1989}  is in fact a LHS model for Werner states.
We only have a few LHS models beyond Werner's original construction, such as the ones in \cite{PRL2014oneway,PRL2005LHVBit,Arxiv2015LHV,JOSAB2015Steering,arxiv2015UnSteer}.

In this work we investigate the construction of  LHS models,
considering arbitrary projective measurements on two-qubit states.
Our work begins with the Bell diagonal states, or, say, \textit{T} states \cite{cen2002local,horodecki1996information,dakic2010necessary}, whose eigenstates are four Bell basis and reduced states maximally mixed.
By using a transformation on the Bloch vectors of the hidden states, we convert this problem to the one for Werner states and obtain the optimal model given by Jevtic \textit{et al.} \cite{JOSAB2015Steering} based on the steering ellipsoid \cite{PRL2014Ellips}.
This transformation enables us to present a criterion sufficient for a two-qubit state to admit a LHS model.
Our criterion is clearly better than the one in \cite{arxiv2015UnSteer}, when  local Bloch vectors tend to zero.
We also compare the maximum allowable lengths of Bloch vectors under the two criteria.

\section{The LHS model and EPR steering}\label{SecLHS}

Let $\rho_{AB}$ denote the entangled quantum state shared by Alice and Bob and  $ \Pi^x_a  $ be Alice's measurement operator of an observable labeled by $x$, corresponding to outcome $a$.
After her measurement, the unnormalized postmeasured state left to Bob is
\begin{eqnarray}\label{Assem}
\rho_{a|x}=\tr_A [(\Pi^x_a \otimes \openone )\rho_{AB}],
\end{eqnarray}
where $\openone $ is the unit operator of Bob's subsystem and $\tr_A$ is the partial trace over Alice's part.
The conditional probability of Alice's outcome is given by $P_A(a|x)=\tr \rho_{a|x}$ and the normalized state prepared for Bob $\tilde{\rho}_{a|x}=\rho_{a|x}/P_A(a|x)$.
The reduced state of Bob satisfies $\rho_{B}=\tr_A  \rho_{AB} =\sum_a\rho_{a|x}$ for all measurements $x$, ensuring that Alice cannot signal to Bob.

A LHS model is defined as
\begin{eqnarray}\label{LHS}
\rho^{\rm LHS}_{a|x}=\int \omega(\lambda)P_A(a|x,\lambda)\rho_{\lambda} d \lambda,
\end{eqnarray}
where $\lambda$ represents a classical (hidden) variable with a distribution $\omega(\lambda)$, $\rho_{\lambda}$ is a state of Bob's system depending on $\lambda$, and $P_A(a|x,\lambda)$ is the probability of outcome $a$ under the condition of $x$ and $\lambda$.
If there exists a LHS model satisfying
\begin{eqnarray}\label{AssEqLHS}
\rho_{a|x}=\rho^{\rm LHS}_{a|x}
\end{eqnarray}
for all the measurements, the results of Alice's measurements can be simulated by a LHS strategy without any entangled state.
Namely, after generating the random variable $\lambda$, Alice prepares a single particle state $\rho_{\lambda}$, sends it to Bob, announces that she measures  $x$, and obtains the outcome $a$ according to the condition probability $P_A(a|x,\lambda)$.
The receiver can not distinguish whether his state and $a$ are the results of Alice's local measurement on $\rho_{AB}$ or she cheats by using the LHS strategy.

The EPR steering from Alice to Bob is demonstrated by the nonexistence of a LHS model satisfying (\ref{AssEqLHS}).
It is the nonlocal correlation, reflected in the effect of Alice's local measurement on Bob's states, which cannot be simulated by a strategy of single-particle-state preparation.
The EPR steering from Bob to Alice also can be defined by reversing their roles in the above.

From the form of $\rho^{\rm LHS}_{a|x}$ in (\ref{LHS}), one can find that  EPR steering is stronger than entanglement and weaker than Bell nonlocality  \cite{PRL2007Steering}.
If $P_A(a|x,\lambda)$ is restricted to conditional probabilities of measurements on Alice's single-particle states,  $\rho^{\rm LHS}_{a|x}$ represents the assemblage of a separable state.
On the other hand, one can derive the joint measurement probability for a state with the LHS model as $P(a,b|x,y)=\int d \lambda \omega(\lambda)P_A(a|x,\lambda)P_B(b|y,\rho_{\lambda})$,
where $P_B(b|y,\rho_{\lambda})=\tr_B (\Pi^y_b \rho_{\lambda} )$, with $\Pi^y_b$ being  Bob's measurement operator of  observable $y$ and outcome $b$.
 Obviously, it is a LHV model with a constraint on the conditional probability of Bob's outcome.

\section{Requirements of two-qubit states}

Under the condition of  preserving steerability (or unsteerability) from Alice to Bob, an arbitrary two-qubit state can always be converted into in the canonical form\cite{arxiv2015UnSteer}
\begin{eqnarray}\label{CanonRho}
\rho_{AB}=\frac{1}{4}\biggr(\openone +\vec{a}\cdot \vec{\sigma}\otimes \openone+  \sum_{i=x,y,z}T_{i}\sigma_{i}\otimes\sigma_{i}\biggr),
\end{eqnarray}
with only one Bloch vector $\vec{a}$ on Alice's side and  a diagonal spin correlation matrix $T=\Diag [T_x,T_y,T_z]$.
That is, it is universal to consider the LHS model of the state (\ref{CanonRho}).

In this work we focus on the case of Alice's local von Neumann measurements, which can be expressed by the projector
\begin{eqnarray}
\Pi^x_a=\frac{1}{2}(\openone+ a \vec{x} \cdot \vec{\sigma})
\end{eqnarray}
 with outcome $a=\pm1$, $\vec{x}$ denoting a unit vector on the Bloch sphere and $\vec{x} \cdot \vec{\sigma}= x_1 \sigma_x +x_2 \sigma_y+x_3 \sigma_z$.
After her measurements, Bob's particle is left in the unnormalized state
\begin{eqnarray}\label{AssemC}
\rho_{a|x}&=&\tr_A [(\Pi^x_a \otimes \openone )\rho_{AB}]\nonumber \\
&=& \frac{1}{4} \bigr[(1+a \vec{a}\cdot \vec{x})\openone +  a (T \vec{x})\cdot \vec{\sigma} \bigr],
\end{eqnarray}
where $T \vec{x} = (T_x x_1, T_y x_2, T_z x_3)^{\mathbf{\top}} $.

Let us look briefly at the LHS models $\rho^{\rm LHS}_{a|x}$, defined in (\ref{LHS}) for a two-qubit system.
It is universal to take the local hidden states $\rho_{\lambda}$ to be pure qubit states, as the eigenvalues of mixed states can be merged into the distribution $\omega(\lambda)$.
Hence, the hidden variable can be represented by the unit Bloch vector $\vec{\lambda}= (\sin\theta \cos \phi,\sin\theta \sin\phi,\cos\theta )$ with \begin{eqnarray}
\rho_{\lambda}=\frac{1}{2}(\openone+   \vec{\lambda} \cdot \vec{\sigma}) .
 \end{eqnarray}
In addition, the integral is over the Bloch sphere with the surface element
$
d \lambda = \sin \theta d \theta d \phi.
$
The conditional probability can be written as
\begin{eqnarray}
P_A(a|x,\lambda)=\frac{1}{2} [1+ a f (x,\lambda)] ,
\end{eqnarray}
where the function $f (x,\lambda) \in [-1,1]$.
Substituting these into the definition of LHS models in (\ref{LHS}) and requiring it to conform to the assemblage in (\ref{AssemC}), one can find the requirements on $\omega(\lambda)$ and  $f (x,\lambda) $ as
\begin{subequations}\label{Reqs}
\begin{align}
&\int \omega(\lambda) d \lambda =1 , \label{omega1}\\
&\int \omega(\lambda) f (x,\lambda) d \lambda = \vec{a}\cdot\vec{x} , \label{a0} \\
&\int \omega(\lambda) \vec{\lambda} d \lambda =0 , \label{b0} \\
&\int \omega(\lambda)  f (x,\lambda) \vec{\lambda} d \lambda = T \vec{x}. \label{Tx}
\end{align}
\end{subequations}
The first relation is the normalization of distribution $\omega(\lambda)$, and the second and third ones correspond to the  local Bloch vectors.
Consequently, constructing a LHS model for  the canonical state (\ref{CanonRho}) is equivalent to finding a solution of $\omega(\lambda)$ and $ f (x,\lambda) $ satisfying these requirements.

\section{Bell diagonal state}

Let us begin with the Bell diagonal state \cite{cen2002local,horodecki1996information,dakic2010necessary}, which is of the canonical form (\ref{CanonRho}) and with  Alice's Bloch vector $\vec{a}=0$.
In this case, the physical region of $T$ is a tetrahedron in the space of $(T_x,T_y,T_z)$,  defined by the set of vertices $(-1, -1, -1)$, $(-1,1,1)$, $(1,-1, 1)$, and $(1, 1, -1)$ corresponding to four Bell basis states.
A separable Bell diagonal state is located in  the octahedron satisfying $\sum_{i=x,y,z}|T_{i}| \leq 1$ \cite{horodecki1996information}.
We may assume that the matrix $T$ is invertible, satisfying $\det T=T_x T_y T_z  \neq 0 $, as a Bell diagonal state  with degenerate  $T$  is separable \cite{PRL2014Ellips}.

Werner states are special Bell diagonal states with $T_x = T_y= T_z <0$.
A solution to the requirements (\ref{Reqs}) for this symmetric situation is given by
\begin{eqnarray}\label{LHSWerner}
\omega(\lambda)=\frac{1}{4 \pi}, \ \ \ f (x,\lambda)= - q \sgn (\vec{\lambda} \cdot \vec{x}),
\end{eqnarray}
where $q\in [0,1]$ and $\sgn $ is the sign function.
Then Eqs. (\ref{omega1})-(\ref{b0}) hold and (\ref{Tx}) is
 \begin{eqnarray}\label{TxWerner}
\int \frac{1}{4 \pi} \bigr[ -q \sgn (\vec{\lambda} \cdot \vec{x}) ] \vec{\lambda} d \lambda = -\frac{1}{2} q \vec{x}.
\end{eqnarray}
The maximum of $q=1$, and simultaneously $|T_i|=1/2$, represents the EPR-steerable boundary for Werner states.
These results can help us solve  the problem of general Bell diagonal states.

Suppose that $T_0$ is a matrix on the EPR-steerable boundary for Bell diagonal states, which can be labeled a family of states with  $T= t T_0$ and $t>0$.
We first give the equation for the boundary by constructing a LHS model for the critical state and then prove the nonexistence of a solution to the requirements (\ref{Reqs}) when $t>1$.
In this sense, the LHS model given below is optimal.
Actually,  it is exactly consistent with the one  given by Jevtic \emph{et al. } \cite{JOSAB2015Steering} and was proved to be optimal by Nguyen and Vu \cite{EPL2016Tstate}.
However, we present an elegant way of generating the model and proving its optimality.
Following our approach, one may generalize the existing results to more general cases.

We start with the relation (\ref{Tx}). Performing  the inverse matrix $T_0^{-1} $ on it,  one obtains
 \begin{eqnarray}\label{Tx1}
\int   \omega(\lambda)  f (x,\lambda) (T_0^{-1} \vec{\lambda})  d \lambda =  t \vec{x}.
\end{eqnarray}
 The integral on the left-hand can be transformed into the one over the unit vector
\begin{eqnarray}
 \vec{\lambda}' =  {|T_0^{-1} \vec{\lambda}|}^{-1} {T_0^{-1} \vec{\lambda}} ,
\end{eqnarray}
where $|T_0^{-1} \vec{\lambda}|$ is the Euclidean vector norm of $T_0^{-1} \vec{\lambda}$.
We also have the relations
\begin{eqnarray}
\vec{\lambda} = {|T_0\vec{\lambda}'|}^{-1} {T_0\vec{\lambda}'},\ \ \
|T_0\vec{\lambda}'| |T_0^{-1} \vec{\lambda}| =1.
\end{eqnarray}
In polar coordinates, $\vec{\lambda}'= (\sin\theta' \cos \phi',\sin\theta' \sin\phi',\cos\theta' )$ the surface element in polar coordinate is $d \lambda' =\sin\theta' d\theta' d \phi'$.
It is connected with the one of  $\vec{\lambda}$ by the Jacobian determinant as
\begin{eqnarray}
d \lambda = |\det T_0| |T_0\vec{\lambda}'|^{-3} d \lambda'.
\end{eqnarray}
The unit vector $\vec{\lambda}'$ is also a hidden variable, with a one-to-one correspondence to $\vec{\lambda}$,
and its distribution $\omega'(\lambda') $ satisfies
 \begin{eqnarray}
\omega(\lambda) d \lambda = \omega'(\lambda')  d \lambda'.
\end{eqnarray}
The relation (\ref{Tx1}) can be rewritten as
\begin{eqnarray}\label{Tx2}
\int \omega'(\lambda')|T_0\vec{\lambda}'|^{-1}  f (x,\lambda)   \vec{\lambda}' d \lambda' =  t \vec{x}.
\end{eqnarray}

We first derive a LHS model for the critical Bell diagonal state.
When $t=1$,  from the integral (\ref{TxWerner}), it is very easy to find a pair of $ \omega'(\lambda') $ and $f (x,\lambda) $ satisfying (\ref{Tx2}) as
\begin{eqnarray}\label{LHSTPrim}
 \omega'(\lambda') = \frac{1}{2 \pi } |T_0\vec{\lambda}'|,\ \ \  f (x,\lambda)=   \sgn(\vec{\lambda}'\cdot\vec{x}).
\end{eqnarray}
That is, the relation (\ref{Tx1}), or equivalently (\ref{Tx}), with $t=1$, is satisfied by
 \begin{eqnarray}\label{LHST}
 \omega(\lambda)= \frac{1}{2 \pi|\det T_0 | |T_0^{-1} \vec{\lambda}|^4},\ \ \
 f (x,\lambda)=   \sgn (\vec{\lambda} \cdot \vec{x}' ),\ \
\end{eqnarray}
where $\vec{x}'  = {|T_0^{-1} \vec{x}|}^{-1} {T_0^{-1} \vec{x}} $ is a unit vector defined similarly to $\vec{\lambda}'$.

Let us substitute them into Eqs. (\ref{omega1})-(\ref{b0}).
The symmetries $ \omega(\lambda)=  \omega(-\lambda)$ and $f (x,-\lambda)=-f (x,\lambda)$ make it is intuitive to confirm the  integrals in both (\ref{a0}) and (\ref{b0})  to be zero.
The normalization condition (\ref{omega1}) leads to the equation of the spin correlation matrix $T_0$ as
 \begin{eqnarray}\label{surface}
2\pi |\det T_0| =  \int  |T_0^{-1} \vec{\lambda}|^{-4} d \lambda.
\end{eqnarray}
In the coordinate system of $\vec{\lambda}'$,  the normalization condition is equivalent to
\begin{eqnarray}
 \int  \frac{1}{2\pi}|T_0 \vec{\lambda}'| d \lambda' =1,
\end{eqnarray}
 which defines the surface of the region for nonsteerable states in the space of $(T_x,T_y,T_z)$.
These coincide the results of Jevtic \emph{et al. } \cite{JOSAB2015Steering}, and  an explicit expression for the integral in (\ref{surface}) can be found in their work.

Next we show the nonexistence of $\omega(\lambda)$ and $f(x,\lambda)$ when $t>1$, by utilizing the results of the Werner state again.
That is, the above LHS model is the optimal one that maximizes the visibility parameter $t$.
The parameter $t$ can be obtained by the dot product between $\vec{x}$ and  Eq. (\ref{Tx1}) as
\begin{eqnarray}\label{Txt}
\int   \omega(\lambda)  f (x,\lambda)  |T_0 \vec{x}'|^{-1} \vec{x}' \cdot \vec{\lambda}  d \lambda =  t.
\end{eqnarray}
Multiplying it by $ |T_0 \vec{x}'|/2\pi$ and integrating  over the sphere of $\vec{x}'$, one has
\begin{eqnarray}\label{Txt1}
\int   \omega(\lambda)  \biggr[ \int \frac{1}{2 \pi} f (x,\lambda) \vec{x}' \cdot \vec{\lambda} d x'  \biggr] d \lambda =  t,
\end{eqnarray}
where $d x'$ is the surface element.
When $f (x,\lambda) =\sgn ( \vec{x}' \cdot \vec{\lambda}) $, the integral in the square brackets reaches its maximum, $1$, which also can be noticed in (\ref{TxWerner}).
Since $ \omega(\lambda)  >0$,  the visibility parameter $t\leq \int   \omega(\lambda)   d \lambda=1$.
Therefore, $t>1$,  or equivalently $\int   |T\vec{\lambda}| d \lambda>2 \pi$, is a necessary and sufficient condition for steerability of a Bell diagonal state.

%%%%%%%%%%%%%%%%%%%%%%%%%%%%%%%%%%%%%%%%%%%%%%%%%%%%%%%%

\section{ Sufficient criterion for unsteerability}

We now turn to the general canonical states with nonzero $\vec{a}$.
Our goal is to derive a sufficient criterion for unsteerability by constructing a LHS model for the assemblage (\ref{AssemC}).

An existing criterion is given by Bowles \emph{et al.} \cite{arxiv2015UnSteer} as
\begin{eqnarray}\label{condition16}
\max_{\vec{x}} [(\vec{a}\cdot \vec{x})^2  -(1-2 t |T_0 \vec{x}|)] \leq 0.
\end{eqnarray}
For a fixed $T_0$,  the constraint specifies the range of $\vec{a}$ and $t$.
This simple condition allows one to detect one-way steering and provides the simplest such examples \cite{arxiv2015UnSteer}.
However, an obvious disadvantage of this result is its deviation from the EPR-steerable boundary for Bell diagonal states when the Bloch vector $\vec{a}$  tends to zero.
This comes from  the choice of distribution for hidden states, which is  uniform over the Bloch sphere.

We take the local hidden Bloch vectors with the distribution $\omega( \lambda)$ in (\ref{LHST}) and define
\begin{eqnarray}\label{fpm}
 f (x,\lambda) \!=\! q(\vec{x})  \sgn (\vec{\lambda} \! \cdot \!  \vec{x}' \! +\!  c ) \!+\! [1\!-\! q(\vec{x})]\sgn (\vec{\lambda} \! \cdot \! \vec{x}' \!-\! c ),
\end{eqnarray}
with $ c, q(\vec{x}) \in[0,1]$.
Then the conditions (\ref{omega1}) and (\ref{b0})  hold, as they depend only on $\omega( \lambda)$.
Substituting  $\omega( \lambda)$ and $ f (x,\lambda)$ into the condition (\ref{Tx}), one has
\begin{eqnarray}
\int \omega(\lambda)  f(x,\lambda) \vec{\lambda} d \lambda =(1-c^2)T_0 \vec{x},
\end{eqnarray}
which is independent of $q(\vec{x})$.
This result can  be integrated easily by utilizing its identical form (\ref{Tx2}).
The calculation details are similar to the ones of the LHS model in \cite{arxiv2015UnSteer}.
That is, our choice of $\omega( \lambda)$ and $ f (x,\lambda)$ fulfills the condition (\ref{Tx}) with  $t=1-c^2$.
Setting $c=\sqrt{1-t}$ and $T=t T_0$, the integral in relation (\ref{a0}) can be expressed as
\begin{eqnarray}
\int \omega(\lambda)  f(x,\lambda) d \lambda =[2 q(\vec{x})-1] Y(T_0,t,\vec{x}),
\end{eqnarray}
where $ Y(T_0,t,\vec{x})>0$, and is equal to the integral $\int (|T_0 \vec{\lambda}'|/2 \pi) d \lambda'$ over the region of $\vec{\lambda}'  \cdot   \vec{x} <\sqrt{1-t}$.
Then we obtain a sufficient criterion for unsteerability of any canonical state as
\begin{eqnarray}\label{conditionT}
\max_{\vec{x}} [(\vec{a}\cdot \vec{x})^2  - Y^2(T_0,t,\vec{x}) ] \leq 0.
\end{eqnarray}
It is equivalent to the existence of $q(\vec{x})$, making our choice of $\omega( \lambda)$ and $ f (x,\lambda)$ fulfill the condition (\ref{a0}).

Obviously,  our construction  eliminates the defect of the inequality (\ref{condition16}) with a short Bloch vector.
Alternatively, one can take the limit of the visibility parameter tending to one.
The condition (\ref{conditionT}) is fulfilled by arbitrary critical Bell diagonal states,
whereas only the Werner state is allowed by the inequality (\ref{condition16}), as the maximal absolute eigenvalue of an anisotropic $T_0$ is larger than $1/2$ \cite{JOSAB2015Steering}.

A natural question is whether our criterion always performs better than the condition (\ref{condition16}).
To compare  them further, we turn to another extreme, where the Bloch vector $\vec{a}$ reaches its maximum length for fixed $t$ and $T_0$.
Since it is a complex problem to perform general maximizations in the two inequalities,  we give the results of  two spacial cases below.

\emph{Case I.} When $\vec{a}$ is an eigenvector of $T_0$ corresponding to the largest  absolute eigenvalue, its maximum length satisfying (\ref{conditionT})  can be proved to be always larger than the one under the condition (\ref{condition16}).
Without loss of generality,  we assume $|T_{0,x}|\leq |T_{0,y}| \leq |T_{0,z}|$ and define  $X(T_0,t,\vec{x})=\sqrt{1-2 t |T_0\vec{x}|}$.
When $\vec{x}=\hat{k}$, $(\vec{a}\cdot \vec{x})^2$ and $ |T_0\vec{x}|$ achieve their maximums, with $\hat{k}$ being a unit vector in the $z$ direction,
while $Y(T_0,t,\vec{x}) $ reaches its minimum, as $Y(T_0,t,\vec{x}) $ is proportional to  an average radius over the region of $\vec{\lambda}'  \cdot   \vec{x} <\sqrt{1-t}$ on the ellipsoid defined by $r=|T_0 \vec{\lambda}'|$.
Hence, the maximizations in both inequalities occur when $\vec{x}=\hat{k}$.
% $\int (|T_0 \vec{\lambda}'|/2 \pi) d \lambda'$, over the region of $\vec{\lambda}'  \cdot   \vec{x} <\sqrt{1-t}$.
%It is intuitive to find that the maximisations in both the inequalities occur when $\vec{x}$ parallel to $\vec{a}$.
When $t=0$, $Y(T_0,0,\hat{k})=X(T_0,0,\hat{k})=1$.
%, with $\hat{k}$ being a unit vector in the $z$ direction.
One can concludes that $Y(T_0,t,\hat{k}) \geq X(T_0,t,\hat{k}) $ when $2t|T_{0,z}|\in [0,1]$, by proving their partial derivatives to satisfy $0\geq \partial Y(T_0,t,\hat{k}) /\partial t \geq \partial X(T_0,t,\hat{k})/\partial t $, the details for which are shown in the Appendix.
This leads to the claim stated of this case.

\begin{figure}
% \begin{flushright}
\includegraphics[width=8.50cm]{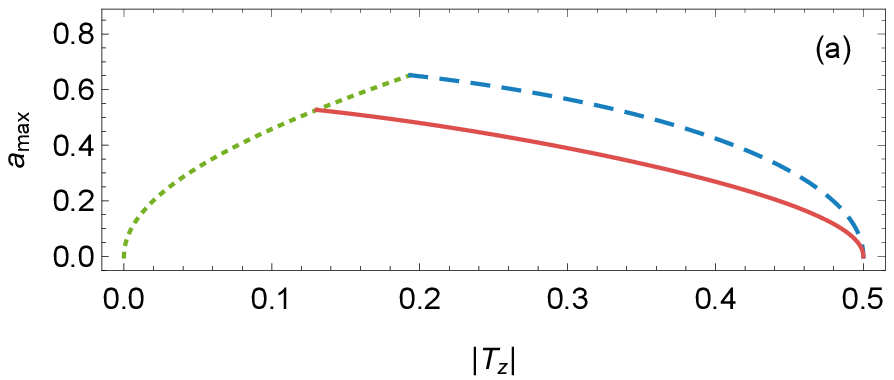}
\\
$ $
\\
\includegraphics[width=8.50cm]{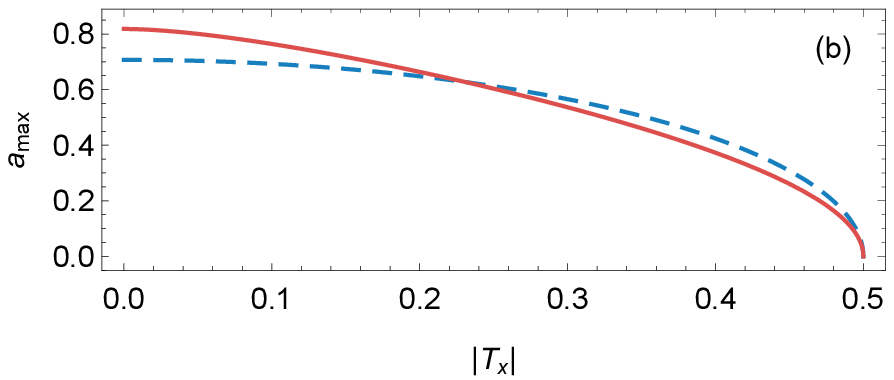}
%\end{flushright}
\caption{
Maximal lengths of $\vec{a}$, (a) along the $z$ axis, as a function of $|T_z|$ when $|T_z|\leq|T_x|=1/2$ and (b) along the $x$ axis, as a function of $|T_x|$ when $|T_z|=1/2$,
allowed by the conditions; the dashed blue lines are for (\ref{condition16}) and the solid red lines are for (\ref{conditionT}).
When the results from these conditions exceed the physical region, the maximum is given by the positive-semidefinite condition for a state and is shown by the dotted green line.
} \label{Amax}
\end{figure}

\emph{Case II.} When $\vec{a}$ is an eigenvector of $T_0$ corresponding to the smallest absolute eigenvalue, each of the two criteria has its own advantages.
First, when $2t  \max\{|T_{0,x}|,|T_{0,y}|,|T_{0,z}|\}>1$ and $t<1$, no state is allowed by the condition (\ref{condition16}), while  the solution set of the  inequality  (\ref{conditionT}) is nonempty as $Y(T_0,t,\vec{x})>0$.
Second, we turn to the region of $2t  \max\{|T_{0,x}|,|T_{0,y}|,|T_{0,z}|\}\leq1$ and consider the matrices $T_0$ with an axial symmetry that $|T_{0,x}|=|T_{0,y}|$.
To display their difference clearly, we choose $t \max\{|T_{0,x}|,|T_{0,z}|\}=1/2$, as the two criteria are equivalent when $T_{0}$ is isotropic.
For fixed $|T_{x}|=|T_{y}|$ and $|T_z|$, a physical Bloch vector satisfies $|\vec{a}| \leq  \min\{1-|T_z|, \sqrt{|T_z| (|T_z|+2)} \}$ when $|T_z|\leq |T_x|=1/2$ and $|\vec{a}| \leq  \sqrt{-|T_x|+3/4 }$ when $|T_x|\leq |T_z|=1/2$.
With these constraints, in Fig. \ref{Amax} we show the maximal lengths of $\vec{a}$ satisfying one of  the two criteria (\ref{condition16})  and   (\ref{conditionT}).
%A positive semidefinite state requires
The condition (\ref{condition16}) performs better than the one  (\ref{conditionT}) when $|T_z|\leq |T_x|=1/2$ or  $0.22\lesssim|T_x|\leq |T_z|=1/2$,
while the latter exceeds the former when $|T_x| \lesssim 0.22 $ and $ |T_z|=1/2$.
Here our numerical result is worth mentioning, namely that the maximization in the condition (\ref{condition16}) occurs at $\vec{x} \perp \vec{a} $, while $\vec{x} \parallel \vec{a} $ for the one in  (\ref{conditionT}) .
This indicates that the dependences on $\vec{x}$ decrease from $2t|T_0 \vec{x}|$ and $(\vec{a}\cdot \vec{x})^2$ to $Y^2(T_0,t,\vec{x})$.

%for a state
%In the region of  the green dotted line in (a), the maximum
 %a  positive semidefinite $\rho_{AB}$
 % derived from the conditions
 %The green dotted line shows the maximum allowed by a  positive semidefinite matrix, where the results derived from inequalities  (\ref{condition16}) and   (\ref{conditionT}) exceed the physical region.

\section{Summary}\label{SecSumm}

We presented a simple approach to generate LHS models for Bell diagonal states from the results of Werner states.
The requirements of Alice's response function and the distribution of hidden states were expressed as four equations.
The key step of our process is mapping the equation corresponding to the correlation matrix to the one of the Werner state.
The latter also enables us to give a concise proof for the optimality of LHS models.

Based on the mapping, we constructed a class of LHS models for canonical states, which led to a sufficient criterion for unsteerability.
Such a criterion becomes a sufficient and necessary one when the Bloch vector vanishes.
In addition, with a Bloch vector along the long axis of the correlation matrix, it can be proved strictly to perform better than the one in \cite{arxiv2015UnSteer}.

Our approach shows the possibility of generating local models from existing results, with a high symmetry, for the cases with a lower symmetry.
This is non-trivial, as it leads to the optimal LHS models for Bell diagonal states and a sufficient criterion for unsteerablity, with some advantages over the existing one in \cite{arxiv2015UnSteer}.

It would be interesting to extend our results in several directions.
On the one hand, one can  try to derive more results for Bell diagonal states by generalizing the mapping to Werner states, such as LHS models for positive-operator-valued measures \cite{PRA2015HiddenSteering} and general LHV models \cite{PhysRevA.73.062105,PHDLHV}.
In addition, whether our idea can be adapted to define Bell diagonal states in higher -dimensional systems is an interesting question.
On the other hand, we may  further  optimize the LHS models for canonical states, at least for some special  cases, e.g., with an axial symmetry.
Several published results may be instructive for this direction, such as the generalization on Werner's distribution \cite{PRL2014oneway} and the geometrical approach to steerability \cite{EPL2016Tstate,PRA2016Geo}.

\begin{acknowledgments}
We thank Q.-H. Yang, L. Zhang, Z.-P. Xu, X.-J. Ye, H.-Y. Su, and J.-L. Chen for discussions.
We are grateful for comments from S. Jevtic, F. Hirsch, and Y.-C. Wu  on an earlier version of the paper.
This work was supported by the NSF of China (Grants No. 11675119, No. 11575125, and No. 11105097).
\end{acknowledgments}

%\bibliography{LHSBellDS}

\begin{appendix}
\begin{widetext}
\section{PROOF OF THE INEQUALITY FOR PARTIAL DERIVATIVES IN CASE I}

In this Appendix we provide details for the relation of the partial derivatives in case I,
\begin{eqnarray}\label{der}
\frac{\partial  } {\partial t } X(T_0,t,\hat{k})  \leq \frac{\partial }{\partial t}  Y(T_0,t,\hat{k}) \leq 0.
\end{eqnarray}

First, we directly obtain
\begin{eqnarray}\label{DX}
\frac{\partial  } {\partial t } X(T_0,t,\hat{k})  =\frac{-|T_{0,z}|}{\sqrt{1-2t |T_{0,z}|}}  \leq \frac{-|T_{0,z}|}{\sqrt{1- t }  }<0,
\end{eqnarray}
as the largest absolute eigenvalue $|T_{0,z}| \geq 1/2$ \cite{JOSAB2015Steering}.
Second, the expression for $Y(T_0,t,\hat{k})$ is given by
\begin{eqnarray}
Y\!\!\!\!\!\!\! && (T_0,t,\hat{k}) \!  = \!  \frac{1}{2\pi}  \int^{\pi - \theta_{t}}_{\theta_{t}} \! \! \!  d \theta \int^{2 \pi}_{0}  \! \! \!  d  \phi   \sin \! \theta   \nonumber \\
\!\!\!\!\!\!\! && \times \sqrt{T_{0,z}^2 \! \cos^2  \! \theta  \!    +   \!     T_{0,x}^2 \! \sin^2  \! \theta \! \cos^2  \! \phi  \!     +  \!     T_{0,y}^2 \! \sin^2  \! \theta \! \sin^2 \! \phi},   \ \ \
\end{eqnarray}
 where $\theta_{t}=\arccos \sqrt{1-t}$.
It can be rewritten as
\begin{eqnarray}
Y(T_0,\!\!\!\!\!\!\! && t,\hat{k}) \!  = \!  \frac{1}{2\pi}  \int^{\sqrt{1-t}}_{-\sqrt{1-t}}  \!  d z \int^{2 \pi}_{0}  \! \! \!  d  \phi     \nonumber \\
\!\!\!\!\!\!\! && \times \sqrt{T_{0,z}^2  z^2  \!  + \! ( T_{0,x}^2 \! \cos^2 \! \phi \! + \! T_{0,y}^2 \! \sin^2 \! \phi)(1-z^2)}, \ \ \
\end{eqnarray}
by defining $z=\cos \theta$. Then the partial derivative satisfies
\begin{eqnarray}\label{DY}
0 \!\!\!\!\!\!\! && >
\frac{\partial }{\partial t}Y(T_0,t,\hat{k}) \!  = \!  \frac{-1}{ \sqrt{1-t}}   \int^{2 \pi}_{0}  \! \! \!  d  \phi  \frac{1}{2\pi}    \nonumber \\
 && \ \ \ \ \ \ \ \ \ \times \sqrt{T_{0,z}^2  (1-t)  \!  + \! ( T_{0,x}^2 \! \cos^2 \! \phi \! + \! T_{0,y}^2 \! \sin^2 \! \phi)t}  \ \ \  \nonumber \\
 \!\!\!\!\!\!\! && \geq \frac{-1}{ \sqrt{1-t}}  |T_{0,z}|.
\end{eqnarray}
Finally, the inequalities (\ref{DX}) and (\ref{DY}) lead to (\ref{der}).

\end{widetext}
\end{appendix}

\bibliography{LHSBellDS}

\begin{thebibliography}{27}
\expandafter\ifx\csname natexlab\endcsname\relax\def\natexlab#1{#1}\fi
\expandafter\ifx\csname bibnamefont\endcsname\relax
  \def\bibnamefont#1{#1}\fi
\expandafter\ifx\csname bibfnamefont\endcsname\relax
  \def\bibfnamefont#1{#1}\fi
\expandafter\ifx\csname citenamefont\endcsname\relax
  \def\citenamefont#1{#1}\fi
\expandafter\ifx\csname url\endcsname\relax
  \def\url#1{\texttt{#1}}\fi
\expandafter\ifx\csname urlprefix\endcsname\relax\def\urlprefix{URL }\fi
\providecommand{\bibinfo}[2]{#2}
\providecommand{\eprint}[2][]{\url{#2}}

\bibitem[{\citenamefont{Nielsen and Chuang}(2000)}]{Book}
\bibinfo{author}{\bibfnamefont{M.~A.} \bibnamefont{Nielsen}} \bibnamefont{and}
  \bibinfo{author}{\bibfnamefont{I.~L.} \bibnamefont{Chuang}},
  \emph{\bibinfo{title}{Quantum Computation and Quantum Information}}
  (\bibinfo{publisher}{Cambridge University Press, Cambridge},
  \bibinfo{year}{2000}).

\bibitem[{\citenamefont{Horodecki et~al.}(2009)\citenamefont{Horodecki,
  Horodecki, Horodecki, and Horodecki}}]{RevModPhys.81.865}
\bibinfo{author}{\bibfnamefont{R.}~\bibnamefont{Horodecki}},
  \bibinfo{author}{\bibfnamefont{P.}~\bibnamefont{Horodecki}},
  \bibinfo{author}{\bibfnamefont{M.}~\bibnamefont{Horodecki}},
  \bibnamefont{and}
  \bibinfo{author}{\bibfnamefont{K.}~\bibnamefont{Horodecki}},
  \bibinfo{journal}{Rev. Mod. Phys.} \textbf{\bibinfo{volume}{81}},
  \bibinfo{pages}{865} (\bibinfo{year}{2009}).

\bibitem[{\citenamefont{Modi et~al.}(2012)\citenamefont{Modi, Brodutch, Cable,
  Paterek, and Vedral}}]{RMP2012Vedral}
\bibinfo{author}{\bibfnamefont{K.}~\bibnamefont{Modi}},
  \bibinfo{author}{\bibfnamefont{A.}~\bibnamefont{Brodutch}},
  \bibinfo{author}{\bibfnamefont{H.}~\bibnamefont{Cable}},
  \bibinfo{author}{\bibfnamefont{T.}~\bibnamefont{Paterek}}, \bibnamefont{and}
  \bibinfo{author}{\bibfnamefont{V.}~\bibnamefont{Vedral}},
  \bibinfo{journal}{Rev. Mod. Phys.} \textbf{\bibinfo{volume}{84}},
  \bibinfo{pages}{1655} (\bibinfo{year}{2012}).

\bibitem[{\citenamefont{Brunner et~al.}(2014)\citenamefont{Brunner, Cavalcanti,
  Pironio, Scarani, and Wehner}}]{RMP2014bell}
\bibinfo{author}{\bibfnamefont{N.}~\bibnamefont{Brunner}},
  \bibinfo{author}{\bibfnamefont{D.}~\bibnamefont{Cavalcanti}},
  \bibinfo{author}{\bibfnamefont{S.}~\bibnamefont{Pironio}},
  \bibinfo{author}{\bibfnamefont{V.}~\bibnamefont{Scarani}}, \bibnamefont{and}
  \bibinfo{author}{\bibfnamefont{S.}~\bibnamefont{Wehner}},
  \bibinfo{journal}{Rev. Mod. Phys.} \textbf{\bibinfo{volume}{86}},
  \bibinfo{pages}{419} (\bibinfo{year}{2014}).

\bibitem[{\citenamefont{Augusiak et~al.}(2014)\citenamefont{Augusiak,
  Demianowicz, and Ac{\'\i}n}}]{JPA2014LHV}
\bibinfo{author}{\bibfnamefont{R.}~\bibnamefont{Augusiak}},
  \bibinfo{author}{\bibfnamefont{M.}~\bibnamefont{Demianowicz}},
  \bibnamefont{and}
  \bibinfo{author}{\bibfnamefont{A.}~\bibnamefont{Ac{\'\i}n}},
  \bibinfo{journal}{J. Phys. A: Math. Theor.} \textbf{\bibinfo{volume}{47}},
  \bibinfo{pages}{424002} (\bibinfo{year}{2014}).

\bibitem[{\citenamefont{Werner}(1989)}]{Werner1989}
\bibinfo{author}{\bibfnamefont{R.~F.} \bibnamefont{Werner}},
  \bibinfo{journal}{Phys. Rev. A} \textbf{\bibinfo{volume}{40}},
  \bibinfo{pages}{4277} (\bibinfo{year}{1989}).

\bibitem[{\citenamefont{Wiseman et~al.}(2007)\citenamefont{Wiseman, Jones, and
  Doherty}}]{PRL2007Steering}
\bibinfo{author}{\bibfnamefont{H.~M.} \bibnamefont{Wiseman}},
  \bibinfo{author}{\bibfnamefont{S.~J.} \bibnamefont{Jones}}, \bibnamefont{and}
  \bibinfo{author}{\bibfnamefont{A.~C.} \bibnamefont{Doherty}},
  \bibinfo{journal}{Phys. Rev. Lett.} \textbf{\bibinfo{volume}{98}},
  \bibinfo{pages}{140402} (\bibinfo{year}{2007}).

\bibitem[{\citenamefont{Schr{\"o}dinger}(1935)}]{steer1935}
\bibinfo{author}{\bibfnamefont{E.}~\bibnamefont{Schr{\"o}dinger}},
  \bibinfo{journal}{Proc. Camb. Phil. Soc.} \textbf{\bibinfo{volume}{31}},
  \bibinfo{pages}{555} (\bibinfo{year}{1935}).

\bibitem[{\citenamefont{Branciard et~al.}(2012)\citenamefont{Branciard,
  Cavalcanti, Walborn, Scarani, and Wiseman}}]{PRA2012OneSDIQKD}
\bibinfo{author}{\bibfnamefont{C.}~\bibnamefont{Branciard}},
  \bibinfo{author}{\bibfnamefont{E.~G.} \bibnamefont{Cavalcanti}},
  \bibinfo{author}{\bibfnamefont{S.~P.} \bibnamefont{Walborn}},
  \bibinfo{author}{\bibfnamefont{V.}~\bibnamefont{Scarani}}, \bibnamefont{and}
  \bibinfo{author}{\bibfnamefont{H.~M.} \bibnamefont{Wiseman}},
  \bibinfo{journal}{Phys. Rev. A} \textbf{\bibinfo{volume}{85}},
  \bibinfo{pages}{010301} (\bibinfo{year}{2012}).

\bibitem[{\citenamefont{Law et~al.}(2014)\citenamefont{Law, Bancal, and
  Scarani}}]{JPA2014quantum}
\bibinfo{author}{\bibfnamefont{Y.~Z.} \bibnamefont{Law}},
  \bibinfo{author}{\bibfnamefont{J.-D.} \bibnamefont{Bancal}},
  \bibnamefont{and} \bibinfo{author}{\bibnamefont{Scarani}},
  \bibinfo{journal}{J. Phys. A: Math. Theor.} \textbf{\bibinfo{volume}{47}},
  \bibinfo{pages}{424028} (\bibinfo{year}{2014}).

\bibitem[{\citenamefont{Saunders et~al.}(2010)\citenamefont{Saunders, Jones,
  Wiseman, and Pryde}}]{NT2010Exp}
\bibinfo{author}{\bibfnamefont{D.~J.} \bibnamefont{Saunders}},
  \bibinfo{author}{\bibfnamefont{S.~J.} \bibnamefont{Jones}},
  \bibinfo{author}{\bibfnamefont{H.~M.} \bibnamefont{Wiseman}},
  \bibnamefont{and} \bibinfo{author}{\bibfnamefont{G.~J.} \bibnamefont{Pryde}},
  \bibinfo{journal}{Nat. Phys.} \textbf{\bibinfo{volume}{6}},
  \bibinfo{pages}{845} (\bibinfo{year}{2010}).

\bibitem[{\citenamefont{Wittmann et~al.}(2012)\citenamefont{Wittmann, Ramelow,
  Steinlechner, Langford, Brunner, Wiseman, Ursin, and
  Zeilinger}}]{wittmann2012loophole}
\bibinfo{author}{\bibfnamefont{B.}~\bibnamefont{Wittmann}},
  \bibinfo{author}{\bibfnamefont{S.}~\bibnamefont{Ramelow}},
  \bibinfo{author}{\bibfnamefont{F.}~\bibnamefont{Steinlechner}},
  \bibinfo{author}{\bibfnamefont{N.~K.} \bibnamefont{Langford}},
  \bibinfo{author}{\bibfnamefont{N.}~\bibnamefont{Brunner}},
  \bibinfo{author}{\bibfnamefont{H.~M.} \bibnamefont{Wiseman}},
  \bibinfo{author}{\bibfnamefont{R.}~\bibnamefont{Ursin}}, \bibnamefont{and}
  \bibinfo{author}{\bibfnamefont{A.}~\bibnamefont{Zeilinger}},
  \bibinfo{journal}{New J. Phys.} \textbf{\bibinfo{volume}{14}},
  \bibinfo{pages}{053030} (\bibinfo{year}{2012}).

\bibitem[{\citenamefont{He et~al.}(2011)\citenamefont{He, Reid, Vaughan, Gross,
  Oberthaler, and Drummond}}]{he2011einstein}
\bibinfo{author}{\bibfnamefont{Q.}~\bibnamefont{He}},
  \bibinfo{author}{\bibfnamefont{M.}~\bibnamefont{Reid}},
  \bibinfo{author}{\bibfnamefont{T.}~\bibnamefont{Vaughan}},
  \bibinfo{author}{\bibfnamefont{C.}~\bibnamefont{Gross}},
  \bibinfo{author}{\bibfnamefont{M.}~\bibnamefont{Oberthaler}},
  \bibnamefont{and} \bibinfo{author}{\bibfnamefont{P.}~\bibnamefont{Drummond}},
  \bibinfo{journal}{Phys. Rev. Lett.} \textbf{\bibinfo{volume}{106}},
  \bibinfo{pages}{120405} (\bibinfo{year}{2011}).

\bibitem[{\citenamefont{Bowles et~al.}(2014)\citenamefont{Bowles, V{\'e}rtesi,
  Quintino, and Brunner}}]{PRL2014oneway}
\bibinfo{author}{\bibfnamefont{J.}~\bibnamefont{Bowles}},
  \bibinfo{author}{\bibfnamefont{T.}~\bibnamefont{V{\'e}rtesi}},
  \bibinfo{author}{\bibfnamefont{M.~T.} \bibnamefont{Quintino}},
  \bibnamefont{and} \bibinfo{author}{\bibfnamefont{N.}~\bibnamefont{Brunner}},
  \bibinfo{journal}{Phys. Rev. Lett.} \textbf{\bibinfo{volume}{112}},
  \bibinfo{pages}{200402} (\bibinfo{year}{2014}).

\bibitem[{\citenamefont{Bowles et~al.}(2015)\citenamefont{Bowles, Hirsch,
  Quintino, and Brunner}}]{PRL2005LHVBit}
\bibinfo{author}{\bibfnamefont{J.}~\bibnamefont{Bowles}},
  \bibinfo{author}{\bibfnamefont{F.}~\bibnamefont{Hirsch}},
  \bibinfo{author}{\bibfnamefont{M.~T.} \bibnamefont{Quintino}},
  \bibnamefont{and} \bibinfo{author}{\bibfnamefont{N.}~\bibnamefont{Brunner}},
  \bibinfo{journal}{Phys. Rev. Lett.} \textbf{\bibinfo{volume}{114}},
  \bibinfo{pages}{120401} (\bibinfo{year}{2015}).

\bibitem[{\citenamefont{Cavalcanti et~al.}(2016)\citenamefont{Cavalcanti,
  Guerini, Rabelo, and Skrzypczyk}}]{Arxiv2015LHV}
\bibinfo{author}{\bibfnamefont{D.}~\bibnamefont{Cavalcanti}},
  \bibinfo{author}{\bibfnamefont{L.}~\bibnamefont{Guerini}},
  \bibinfo{author}{\bibfnamefont{R.}~\bibnamefont{Rabelo}}, \bibnamefont{and}
  \bibinfo{author}{\bibfnamefont{P.}~\bibnamefont{Skrzypczyk}},
  \bibinfo{journal}{Phys. Rev. Lett.} \textbf{\bibinfo{volume}{117}},
  \bibinfo{pages}{190401} (\bibinfo{year}{2016}).

\bibitem[{\citenamefont{Jevtic et~al.}(2015)\citenamefont{Jevtic, Hall,
  Anderson, Zwierz, and Wiseman}}]{JOSAB2015Steering}
\bibinfo{author}{\bibfnamefont{S.}~\bibnamefont{Jevtic}},
  \bibinfo{author}{\bibfnamefont{M.~J.} \bibnamefont{Hall}},
  \bibinfo{author}{\bibfnamefont{M.~R.} \bibnamefont{Anderson}},
  \bibinfo{author}{\bibfnamefont{M.}~\bibnamefont{Zwierz}}, \bibnamefont{and}
  \bibinfo{author}{\bibfnamefont{H.~M.} \bibnamefont{Wiseman}},
  \bibinfo{journal}{J. Opt. Soc. Am. B} \textbf{\bibinfo{volume}{32}},
  \bibinfo{pages}{A40} (\bibinfo{year}{2015}).

\bibitem[{\citenamefont{Bowles et~al.}(2016)\citenamefont{Bowles, Hirsch,
  Quintino, and Brunner}}]{arxiv2015UnSteer}
\bibinfo{author}{\bibfnamefont{J.}~\bibnamefont{Bowles}},
  \bibinfo{author}{\bibfnamefont{F.}~\bibnamefont{Hirsch}},
  \bibinfo{author}{\bibfnamefont{M.~T.} \bibnamefont{Quintino}},
  \bibnamefont{and} \bibinfo{author}{\bibfnamefont{N.}~\bibnamefont{Brunner}},
  \bibinfo{journal}{Phys. Rev. A} \textbf{\bibinfo{volume}{93}},
  \bibinfo{pages}{022121} (\bibinfo{year}{2016}).

\bibitem[{\citenamefont{Cen et~al.}(2002)\citenamefont{Cen, Wu, Yang, and
  An}}]{cen2002local}
\bibinfo{author}{\bibfnamefont{L.-X.} \bibnamefont{Cen}},
  \bibinfo{author}{\bibfnamefont{N.-J.} \bibnamefont{Wu}},
  \bibinfo{author}{\bibfnamefont{F.-H.} \bibnamefont{Yang}}, \bibnamefont{and}
  \bibinfo{author}{\bibfnamefont{J.-H.} \bibnamefont{An}},
  \bibinfo{journal}{Phys. Rev. A} \textbf{\bibinfo{volume}{65}},
  \bibinfo{pages}{052318} (\bibinfo{year}{2002}).

\bibitem[{\citenamefont{Horodecki and
  Horodecki}(1996)}]{horodecki1996information}
\bibinfo{author}{\bibfnamefont{R.}~\bibnamefont{Horodecki}} \bibnamefont{and}
  \bibinfo{author}{\bibfnamefont{M.}~\bibnamefont{Horodecki}},
  \bibinfo{journal}{Phys. Rev. A} \textbf{\bibinfo{volume}{54}},
  \bibinfo{pages}{1838} (\bibinfo{year}{1996}).

\bibitem[{\citenamefont{Daki{\'c} et~al.}(2010)\citenamefont{Daki{\'c}, Vedral,
  and Brukner}}]{dakic2010necessary}
\bibinfo{author}{\bibfnamefont{B.}~\bibnamefont{Daki{\'c}}},
  \bibinfo{author}{\bibfnamefont{V.}~\bibnamefont{Vedral}}, \bibnamefont{and}
  \bibinfo{author}{\bibfnamefont{{\v{C}}.}~\bibnamefont{Brukner}},
  \bibinfo{journal}{Phys. Rev. Lett.} \textbf{\bibinfo{volume}{105}},
  \bibinfo{pages}{190502} (\bibinfo{year}{2010}).

\bibitem[{\citenamefont{Jevtic et~al.}(2014)\citenamefont{Jevtic, Pusey,
  Jennings, and Rudolph}}]{PRL2014Ellips}
\bibinfo{author}{\bibfnamefont{S.}~\bibnamefont{Jevtic}},
  \bibinfo{author}{\bibfnamefont{M.}~\bibnamefont{Pusey}},
  \bibinfo{author}{\bibfnamefont{D.}~\bibnamefont{Jennings}}, \bibnamefont{and}
  \bibinfo{author}{\bibfnamefont{T.}~\bibnamefont{Rudolph}},
  \bibinfo{journal}{Phys. Rev. Lett.} \textbf{\bibinfo{volume}{113}},
  \bibinfo{pages}{020402} (\bibinfo{year}{2014}).

\bibitem[{\citenamefont{Nguyen and Vu}(2016{\natexlab{a}})}]{EPL2016Tstate}
\bibinfo{author}{\bibfnamefont{H.~C.} \bibnamefont{Nguyen}} \bibnamefont{and}
  \bibinfo{author}{\bibfnamefont{T.}~\bibnamefont{Vu}}, \bibinfo{journal}{EPL
  (Europhysics Letters)} \textbf{\bibinfo{volume}{115}}, \bibinfo{pages}{10003}
  (\bibinfo{year}{2016}{\natexlab{a}}).

\bibitem[{\citenamefont{Quintino et~al.}(2015)\citenamefont{Quintino,
  V{\'e}rtesi, Cavalcanti, Augusiak, Demianowicz, Ac{\'\i}n, and
  Brunner}}]{PRA2015HiddenSteering}
\bibinfo{author}{\bibfnamefont{M.~T.} \bibnamefont{Quintino}},
  \bibinfo{author}{\bibfnamefont{T.}~\bibnamefont{V{\'e}rtesi}},
  \bibinfo{author}{\bibfnamefont{D.}~\bibnamefont{Cavalcanti}},
  \bibinfo{author}{\bibfnamefont{R.}~\bibnamefont{Augusiak}},
  \bibinfo{author}{\bibfnamefont{M.}~\bibnamefont{Demianowicz}},
  \bibinfo{author}{\bibfnamefont{A.}~\bibnamefont{Ac{\'\i}n}},
  \bibnamefont{and} \bibinfo{author}{\bibfnamefont{N.}~\bibnamefont{Brunner}},
  \bibinfo{journal}{Phys. Rev. A} \textbf{\bibinfo{volume}{92}},
  \bibinfo{pages}{032107} (\bibinfo{year}{2015}).

\bibitem[{\citenamefont{Ac\'\i{}n et~al.}(2006)\citenamefont{Ac\'\i{}n, Gisin,
  and Toner}}]{PhysRevA.73.062105}
\bibinfo{author}{\bibfnamefont{A.}~\bibnamefont{Ac\'\i{}n}},
  \bibinfo{author}{\bibfnamefont{N.}~\bibnamefont{Gisin}}, \bibnamefont{and}
  \bibinfo{author}{\bibfnamefont{B.}~\bibnamefont{Toner}},
  \bibinfo{journal}{Phys. Rev. A} \textbf{\bibinfo{volume}{73}},
  \bibinfo{pages}{062105} (\bibinfo{year}{2006}).

\bibitem[{\citenamefont{Toner}(2007)}]{PHDLHV}
\bibinfo{author}{\bibfnamefont{B.~F.} \bibnamefont{Toner}}, Ph.D. thesis,
  \bibinfo{school}{California Institute of Technology} (\bibinfo{year}{2007}).

\bibitem[{\citenamefont{Nguyen and Vu}(2016{\natexlab{b}})}]{PRA2016Geo}
\bibinfo{author}{\bibfnamefont{H.~C.} \bibnamefont{Nguyen}} \bibnamefont{and}
  \bibinfo{author}{\bibfnamefont{T.}~\bibnamefont{Vu}}, \bibinfo{journal}{Phys.
  Rev. A} \textbf{\bibinfo{volume}{94}}, \bibinfo{pages}{012114}
  (\bibinfo{year}{2016}{\natexlab{b}}).

\end{thebibliography}

\end{document}